\documentclass[a4paper, 11pt]{article}
\usepackage[utf8]{inputenc}
\usepackage{indentfirst}
\usepackage{geometry}
\usepackage{textcomp}
\usepackage{amsmath}
\usepackage{amssymb}
\usepackage{graphicx}
\usepackage{color,xcolor}
\usepackage{mathrsfs}
\usepackage{caption}
\usepackage{dsfont}
\usepackage[colorlinks=true, allcolors=blue]{hyperref}
\usepackage{tcolorbox}
\usepackage{tikz}
\usepackage{tikz-cd}
\usepackage{physics}
\usepackage{diagbox}
\usepackage{listings}
\usepackage{appendix}
\usepackage{array}
\usepackage{subfig}

\catcode`\>=\active \def>{
\fontencoding{T1}\selectfont\symbol{62}\fontencoding{\encodingdefault}}

\newcommand{\tmop}[1]{\ensuremath{\operatorname{#1}}}

\usepackage{amsmath}
\numberwithin{equation}{section}
\usepackage[normalem]{ulem}
\usepackage{color}
\definecolor{darkgreen}{RGB}{40,150,60}

\def\a{\alpha}

\def\d{\delta}

\def\f{\frac}

\def\lm{\lambda}

\def\L{\Lambda}
\def\m{\mu} 
\def\n{\nu} 
\def\nn{\nonumber}
\def\pl{\partial}
\def\p{\phi} 
\def\td{\tilde}

\def\t{\theta}

\def \md{\mathds}

\def\be{\begin{equation}}
\def\ee{\end{equation}}
\def\bag{\begin{aligned}}
	\def\eag{\end{aligned}}
\def\bea{\begin{eqnarray}}
\def\eea{\end{eqnarray}}
\def\ba{\begin{array}}
	\def\ea{\end{array}}

\title{On emergent conformal symmetry near the photon ring}
\author{Bin Chen$^{1,2,3}$,  Yehui Hou$^1$, and Zezhou Hu$^1$}
\date{\today}

\begin{document}

\maketitle
\begin{center}
	{\it
		$^{1}$School of Physics, Peking University, No.5 Yiheyuan Rd, Beijing 100871, P.~R.~China\\
		\vspace{2mm}
		$^{2}$Collaborative Innovation Center of Quantum Matter, No.5 Yiheyuan Rd, Beijing 100871, P.~R.~China\\
		$^{3}$Center for High Energy Physics, Peking University, No.5 Yiheyuan Rd, Beijing 100871, P.~R.~China\\
	}
	\vspace{10mm}
\end{center}

\begin{abstract}
    \vspace{5mm}
   In this note we revisit the emergent conformal symmetry in the near-ring region of warped spacetime. In particular, we propose a novel construction of the emergent near-ring $sl(2,\md{R})_{\text{QNM}}$ symmetry. We show that each eikonal QNM family falls into one highest-weight representation of this algebra,  and $sl(2,\md{R})_{\text{QNM}}$ can be related to the  near-ring isometry group $sl(2,\md{R})_{\text{ISO}}$ in a simple way. Furthermore we find that the coherent state space of $sl(2,\md{R})_{\text{QNM}}$ can be identified with the phase space of the photon ring. \end{abstract}
\newpage

\section{Introduction}

The conformal symmetry plays an important role in studying black hole physics. In AdS$_3$/CFT$_2$ correspondence, the entropy of the BTZ black hole could be computed microscopically from the asymptotic growth of highly excited states in dual CFT$_2$\cite{Strominger:1997eq}. Moreover, the frequencies of the quasi-normal modes of the BTZ black hole appear exactly as the poles in the retarded Green function of the perturbations in the dual CFT$_2$\cite{Birmingham:2001pj,Birmingham:2002ph}. These two correspondences  are essential in the subsequent investigations on the holographic pictures of higher-dimensional black holes\cite{Guica:2008mu,Bredberg:2009pv} (see a nice review\cite{Compere:2012jk}) and warped black holes\cite{Moussa:2003fc,Anninos:2008fx,Chen:2009hg,Chen:2009cg,Chen:2010qm,Song:2011sr}. Especially, for the higher dimensional black holes, the conformal symmetry does not present itself explicitly. In the case of the Kerr black hole, the symmetry either appears only in the isometry of the near horizon geometry of extreme Kerr (NHEK) \cite{Bardeen:1999px} or emerges as the hidden conformal symmetry in the low frequency scattering\cite{Bredberg:2009pv,Castro:2010fd,Chen:2010fr}.

In the past few years, following the release of the black hole images taken by the Event Horizon Telescope (EHT) collaborations\cite{Akiyama:2019cqa}, the physics encoded in the images has been intensely studied. Of particular interest is the black hole photon ring, which has a subring structure. This subring structure displays universal properties of the black holes, which are insensitive to the environment around the black holes\cite{Gralla:2019xty,Johnson:2019ljv,Gralla:2020yvo,Gralla:2020srx,Himwich:2020msm}. The photon ring is produced by the unstably bound orbits of photons. Under a perturbation, the photons on the orbit could escape to infinity after orbiting many times. It turns out that these null trajectories provide eikonal approximations to the quasi-normal modes (QNMs)\cite{Ferrari:1984zz,Mashhoon:1985cya,Cardoso:2008bp,Yang:2012he,Li:2021zct}.

The idea of holography of the photon ring was proposed in \cite{Hadar:2022xag} very recently. On the one hand, it was shown that in the near-ring region of both Schwarzschild and Kerr black holes, there emerges a hidden conformal symmetry, referred to as  $sl(2,\md{R})_{\text{PR}}$, acting on the photon trajectories. More precisely, in the (reduced) phase space of null geodesics, $(T,\Psi,H,L)$ for a Schwarzschild black hole or $(T,\Phi,\Psi,H,L,Q+L^2)$ for a Kerr black hole, the $sl(2,\md{R})_{\text{PR}}$ algebra can be constructed via Poisson bracket. On the other hand, considering the massless wave propagation in the near-ring region, the wave equation reduces to the Schrödinger equation with an upside-down harmonic potential, which has a well-known conformal symmetry $sl(2,\md{R})_{\text{QNM}}$. It turned out that 
the $sl(2,\md{R})_{\text{QNM}}$ algebra could be constructed by Heisenberg algebra, and the quasi-normal modes (QNMs) in the near-ring region can be described in a Fock space\cite{Raffaelli:2021gzh,Hadar:2022xag}. The eikonal 
QNMs fall into the highest-weight representations of $sl(2,\md{R})_{\text{QNM}}$, similar to the cases of the BTZ black hole\cite{Birmingham:2001pj} and the warped black holes\cite{Chen:2010ik,Chen:2011dc,Chen:2010sn}. However, it is perplexing that the entire QNMs constructed in \cite{Hadar:2022xag} are decomposed into \textit{two} highest-weight representations of $sl(2,\md{R})_{\text{QNM}}$, with one tower of descendants having even overtone numbers and the other having odd overtone numbers. 

In a more recent work, the study was generalized to the self-dual warped AdS$_3$ geometry\cite{Kapec:2022dvc}, which is similar to the NHEK geometry and has an exact $sl(2,\md{R})_{\text{ISO}}$ isometry group. It was  showed that the $sl(2,\md{R})_{\text{PR}}$ algebra in the phase space can be induced from $sl(2,\md{R})_{\text{ISO}}$ by simply contracting the Killing vectors with the momentum of the photon. The entire eikonal QNM spectrum  can be constructed by using the $sl(2,\md{R})_{\text{ISO}}$ symmetry as well. However, because each QNM overtone family has two highest-weight representations of $sl(2,\md{R})_{\text{QNM}}$ in the near-ring region, the $sl(2,\md{R})_{\text{QNM}}$ symmetry constructed from the Heisenberg algebra failed to match the $sl(2,\md{R})_{\text{ISO}}$ symmetry.

In this note, we present a novel way to construct $sl(2,\md{R})_{\text{QNM}}$ in the near-ring region. For simplicity, our construction focuses on the self-dual warped AdS$_3$ geometry, but it can be applied to the near-ring regions of four-dimensional black holes as well. We show that within our construction the entire eikonal QNMs spectrum is described by only \textit{one} highest-weight  representation and consequently the newly constructed $sl(2,\md{R})_{\text{QNM}}$ match with the $sl(2,\md{R})_{\text{ISO}}$ symmetry nicely.  Moreover, we show that there is an exact correspondence between the near-ring QNMs and the points in the phase space of null geodesics, which may allow us to further identify the trajectories in the phase space with the coherent states of  $sl(2,\md{R})_{\text{QNM}}$.



\section{Constructions of $sl(2,\md{R})_{\text{PR}}$}\label{PR}

For simplicity, we  show our construction in a self-dual warped AdS$_3$ geometry. The metric,  which can be obtained from the near-horizon approximation of a near-extreme warped AdS$_3$ black hole \cite{Bouchareb:2007yx},  is of the form
\be
\label{AdS}
d s^2=l^2\left[\frac{-d T^2+d x^2}{\sinh^2 x}+\Lambda^2\left(d\phi+\frac{d T}{\tanh x}\right)^2\right]\,,
\ee
where $l$ is the (warped) AdS$_3$ radius, $T$ and $x$ are the time and radial coordinates, respectively. The parameter $\L$ is a warping factor.

The null geodesics in the above spacetime are described by the entire phase space (EPS) with the coordinates $(x^{\m},p_{\m})$, $x^\m = T,x,\p$, $p_x = p_T, p_x, p_\p$, and a constrained Hamiltonian, $\mathcal{H} = \f{1}{2}g^{\m\n}p_{\m}p_{\n}=0$. The Hamilton equations are 
\be
\{x^{\m},\mathcal{H}\}_{\text{EPS}}=\frac{d x^{\m}}{dt}\,, \quad \{p_{\m},\mathcal{H}\}_{\text{EPS}}=\frac{dp_{\m}}{dt}\,,
\ee
where $t$ is the affine parameter, and the Poisson bracket is defined as $\{ A, B \}_{\text{EPS}} = \f{\pl A}{\pl x^\m}\f{\pl B}{\pl p_\m} - \f{\pl B}{\pl x^\m}\f{\pl A}{\pl p_\m}$. Since the metric does not depend on $T$ and $\p$, there are two constants of motion, i.e., the energy and angular momentum, $H = -p_T$, $L=p_{\p}$. Then, the equations of motion are
\bea\label{eom1}
\big(\f{l}{\sinh{x}}\big)^2\frac{dx}{dt} &= &\pm_{x}H\sqrt{V(x)}\,, \nn \\
l^2 \frac{dT}{dt} &= &H\sinh{x}(\sinh{x}+\lm\cosh{x} )\, , \,\nn \\
l^2 \frac{d\p}{dt}& =& H (\lm \Lambda^{-2}-\lm\cosh^2{x}-\sinh{x}\cosh{x} ) \, ,
\eea
where $\pm_x$ denotes the sign of $dx/dt$, and the radial potential reads
\be\label{VV}
V(x) = (1+\lm\coth{x})^2-\f{\lm^2}{\L^2\sinh^2{x}} \ge 0\,.
\ee
Note that we have introduced a dimensionless impact parameter, $\lm = L/H$. The bound photon orbits are determined by
\be\label{VV0}
V(x) = V'(x) = 0\, ,
\ee
which gives the double roots of the potential corresponding to the unstable bound photon orbits. After solving Eq.~\eqref{VV0},  we get
\be\label{bpo1}
\hat{x}_{\pm} = \pm \tanh^{-1}{\sqrt{1-\f{1}{\L^2}}}\,, \quad \hat{\lm}_{\pm} = \mp\f{1}{\sqrt{1-\f{1}{\L^2}}}\,. 
\ee
We assume $\L$ to be greater than $1$ to ensure the existence of bound photon orbits. From now on, we only consider the outer orbit labeled by ``$+$'' and write $\hat{\lm} = \hat{\lm}_+$, $\hat{x} = \hat{x}_+$ for simplicity.

The null geodesics can also be studied in a reduced phase space (RPS) with the coordinates $(x,\p,p_x,p_\p)$, the symplectic form $dp_x \wedge dx + dp_\p\wedge d\p$ and the reduced Hamiltonian $H = -p_T$. From the first two equations in Eq.~\eqref{eom1}, we find that $T = T(x,p_x,p_\p)$ is canonically conjugate to $H$, that is, $\{T,H\}_{\text{RPS}}=1$. Thus, we choose $T$ as the time function. Since we only concentrate on the RPS in the following, we omit the subscript RPS in the Poisson bracket for simplicity.

In addition, we can  simplify the system further by introducing a new variable \cite{Kapec:2022dvc}
\be
\Psi(x,\p,p_x,p_\p) = \int d\p+H\int \bigg[ \coth{x}(1+\lm\coth{x})-\f{\lm}{\L^2\sinh^2{x}} \bigg]\f{dx}{\sqrt{V(x)}}\,. 
\ee
Then, the RPS can be described by the four variables $(T,\Psi,H,L)$ with the symplectic form $dH \wedge dT + dL\wedge d\Psi$. The equations of motion can be rewritten as
\be
\dot{T} = \{T,H\}=1\,, \quad  \dot{\Psi} = \{\Psi,H\}=0\,, \quad
\dot{L} = \{L,H\}=0\,,
\ee 
where the ``dot'' represents the derivative with respect to $T$. Now we study the properties related to the bound photon  orbit in the RPS described by $(T,\Psi,H,L)$. To characterize the deviation from the bound photon orbit, we define
\be
\hat{H} = H - L/\hat{\lm}\,.
\ee
The bound photon orbit has $\hat{H} = 0$. As in \cite{Hadar:2022xag,Kapec:2022dvc}, the relation $\{T,H\}=\{T,\hat{H}\}=1$ can be used to construct a $sl(2,\md{R})$ algebra, which is generated by 
\be
H_{+}=\hat{H}\,, \quad
H_0=-\hat{H} T\,, \quad
H_{-}=\hat{H} T^2\,,
\ee
with the commutation relation
\be
\{H_p,H_q\}=(p-q)H_{p+q}\, , \qquad (p,q=+,0,-) \,.
\ee
Here, the quantity of physical interest is $T\hat{H}$. To see this, we define the finite action of $A$ on $B$ as
\bea
e^{-\a A} B e^{\a A} \equiv  \sum_n \frac{(- \alpha)^n}{n!}
\tmop{ad}_{A}^n B  \, , \quad
\tmop{ad}_{A} B \equiv  \{A, B \}\,,
\eea
with both $A, B$ being the functions in the RPS. Then, the action of $T\hat{H}$ on $\hat{H}$ is
\be\label{choose1}
e^{-\a T\hat{H}} \hat{H} e^{\a T\hat{H}} = e^{-\a} \hat{H}\,,
\ee
which means that $e^{-\a H_0}$ is a finite dilation, leading to the attraction towards the bound photon  orbit if $\a>0$. 

However, the construction of the $sl(2,\md{R})$ algebra is not unique. Based on the relation $\{T,\hat{H}\}=1$, there exist the following constructions of $sl(2,\md{R})$ algebra as well, 
\bea\label{prconstru1}
H_{+}=T\,, \quad
H_0=T \hat{H}\,, \quad
H_{-}=T \hat{H}^2\,,
\eea
or
\bea\label{prconstru2}
H_{+}=\f{1}{2}T^2\,, \quad
H_0=\f{1}{2}T \hat{H}\,, \quad
H_{-}=\f{1}{2} \hat{H}^2\,, 
\eea
and the automorphism of  $sl(2,\md{R})$
\bea
H_{\pm}\rightarrow H_{\mp} \,, \
H_0\rightarrow -H_0\,.
\eea
Note that all the constructions above have $H_0 \propto T\hat{H}$.  Moreover, all the constructions satisfy the same constraint
\begin{equation} \label{constraint1}
H_0^2=H_{+} H_{-}\,.
\end{equation}
Since the algebra commute with $L$, a superselection sectors of the RPS with fixed $L$ (sub-RPS) can be described by three generators $(H_{+},H_{0},H_{-})$ with the constraint (\ref{constraint1}). 

In addition, the relation $\{ \Psi, L \} = 1$ also leads to similar constructions of $sl(2,\md{R})$ algebra. For example, there is  
\be
H_{+}=\hat{L},  \, \
H_0=-\hat{L} \Psi,   \, \
H_{-}=\hat{L} \Psi^2
\ee
with $\hat{L} = L - \td{\lm}H = -\td{\lm}\hat{H}$ describing the deviation from the bound photon orbit, which has $\hat{L} = 0$. 
Here, the dilation operator is $\Psi\hat{L}$, since under a finite action, we have
\be\label{choose2}
e^{-\a \Psi\hat{L}} \hat{L} e^{\a \Psi\hat{L}} = e^{-\a} \hat{L}\,.
\ee
The reason for the existence of two schemes, Eq~\eqref{choose1} and Eq~\eqref{choose2}, is that the bound photon orbit is parameterized only by the impact parameter, $\lm = L/H$. 

\section{Novel construction of $sl(2,\md{R})_{\text{QNM}}$}

In this section, we discuss the properties of the near-ring QNMs introduced in \cite{Kapec:2022dvc}, and present a novel construction of the $sl(2,\md{R})_{\text{QNM}}$ algebra of the QNM spectrum. Although the construction present here is based on the warped AdS$_3$ spacetime, it depends solely on the photon ring and the near-ring region. Thus, our construction can be extended to the four-dimensional Schwarzschild and Kerr black holes straightforwardly.

The isometry group of the warped AdS$_3$ spacetime Eq.~\eqref{AdS} is generated by the following Killing vectors
\begin{equation}\label{Killing}
\begin{aligned}
&\mathcal{L}_0=-\partial_T\,, \\
&\mathcal{L}_{\pm}=e^{\pm T}(-\cosh x \partial_T \mp \sinh x \partial_x+\sinh x \partial_\phi)\,, \\
&\mathcal{W}_0=\partial_\phi\,.
\end{aligned}
\end{equation}
The first three vectors generate an $sl(2,\md{R})$ algebra, referred to as $sl(2,\md{R})_{\text{ISO}}$. The commutation relation is
\begin{equation}
[\mathcal{L}_p,\mathcal{L}_q]=(p-q)\mathcal{L}_{p+q}\, , \qquad (p,q=+,0,-) \,.
\end{equation}
We consider a free scalar field $\Phi(T,x,\phi)$ minimally coupled with gravity. The equation of motion reads
\be
\label{EOM1}
\nabla^2 \Phi=-\frac{1}{l^2}\left[-\mathcal{L}_0^2+\frac{\mathcal{L}_+ \mathcal{L}_- +\mathcal{L}_- \mathcal{L}_+}{2}+\left(1-\frac{1}{\Lambda^2}\right)\mathcal{W}_0^2\right]\Phi=0\,,
\ee
where the Laplacian is rewritten in terms of the Casimir and  commutes with the Lie-derivative along the vector fields defined in Eq.~\eqref{Killing}. Hence, the isometric symmetry is retained in the solution space of $\Phi$. As the translations along $T$ and $\phi$ are obvious Killing vectors, the eigenfunctions of Eq.~\eqref{EOM1} can be labeled by the eigenvalues of the operators 
$\mathcal{L}_0$ and $\mathcal{W}_0$,   
\be
\Phi(T,\phi,x) = e^{im\phi}e^{(-i\hat{\omega}_R+\hat{\omega}_I)T} \psi(x)\,, 
\ee
where $m = 0, \pm 1, \pm 2...$ is the angular quantum number, and $\hat{\omega}=\hat{\omega}_R+i \hat{\omega}_I$ is the complex frequency of the modes, with $\hat{\omega}_R$ and $\hat{\omega}_I$ the real and imaginary parts, respectively.

The QNMs require the ingoing condition at the horizon where $x \rightarrow +\infty$, and the outgoing condition at infinity where $x \rightarrow 0$.  As shown in \cite{Kapec:2022dvc}, the conditions are satisfied by the highest-weight representation of $sl(2,\md{R})_{\text{ISO}}$,
\be\label{hwr1}
\Phi_{m,n}(T,\phi,x) = e^{im\phi}e^{-(i\hat{\omega}_R+\f{1}{2}+n)T} \psi_{m, n}(x)\,, 
\ee
where $n = 0,1,2...$ is the overtone number, and 
\be
\hat{\omega}_R = -m \sqrt{1-\f{1}{\L^2}-\f{1}{4m^2}}\, .
\ee 
The fundamental mode $\Phi_{m,0}$ is annihilated by $\mathcal{L}_+$, i.e., $\mathcal{L}_+ \Phi_{m,0}=0$, which means that it is the highest-weight state,
\be
\mathcal{L}_0 \Phi_{m,0}=\hat{h}\Phi_{m,0}=\bigg(\f{1}{2}+i\hat{\omega}_R\bigg)\Phi_{m,0}\,.
\ee
The descendants $\Phi_{m,n}(n> 0)$ can be obtained by acting $\mathcal{L}_-$ on $\Phi_{m,0}$ successively,
\be
\Phi_{m,n}\propto \mathcal{L}_-^n \Phi_{m,0}\,.
\ee
We would not like to discuss the QNMs for generic $m$, so we do not need the specific form of $\psi_{m,n}(x)$, which were studied in detail in the section 2.3 of \cite{Kapec:2022dvc}.
	From now on, we consider the eikonal limit of the QNMs with $m\gg1$. The eikonal QNMs can be obtained by taking $|m|\rightarrow \infty$ in \eqref{hwr1}, which leads to a simple form of the wave-function
\begin{equation}
\label{eikonal}
\Phi_{m,n}=e^{-(n+\hat{h})T}e^{i m \phi+ i m x}(\sinh x)^{\hat{h}}(\cosh x+\hat{\lambda} \sinh x)^n \,,
\end{equation}
where the highest weight is
\be
\hat{h} = 1/2 + i\hat{\omega}_R\,, \quad \hat{\omega}_R \approx m\hat{\Omega} \,
\ee
with $\hat{\Omega} =\hat{\lm}^{-1}= -\sqrt{1-\Lambda^{-2}}$ being the angular velocity of the (outer) bound photon orbit. 

Next, we focus on the ``near-ring region'' of the eikonal QNMs, defined by
\begin{equation}
\begin{aligned}
\label{nr1}
&|\delta x|\ll1\,, \\
&|m/\hat{\omega}_R-\hat{\lambda}|\ll1\,, \\
&|\hat{\omega}_R| \gg 1\,,
\end{aligned}
\end{equation}
where $\d x=x-\hat{x}_+$ describing the deviation from the bound photon orbit.
In the warped AdS$_3$ spacetime, the last two conditions  indicate the eikonal condition $|m| \gg 1$. Note that in the near-ring region, the eikonal limit holds only if $ |\hat{\omega}_R| \delta x^2 \gg 1$, which is the requirement that in this small region, the wavefront is still rapidly changing \cite{Schutz:1985km}. In this case, the eikonal QNMs in Eq.~\eqref{eikonal} become
\begin{equation}
\label{nearring}
\Phi^{\text{(NR)}}_{m, n}\approx c_0 e^{-(n+\hat{h})T}e^{i m \phi}\left(-\frac{\delta x}{\sinh \tilde{x}}\right)^n  e^{-\frac{i}{2}m \hat{\Omega}k \delta x^2}\,,
\end{equation}
where $k = (\L^2-1)^{-1}$, $c_0 = e^{im\td{x}}(\sinh{\td{x}})^{im\hat{\Omega}}$. The superscript  ``NR'' denotes the near-ring approximation. To check that the $sl(2,\md{R})_{\text{ISO}}$ still exists under such approximation, we act the Lie-derivatives on the QNMs in Eq.~\eqref{nearring} and find
\begin{equation}
\begin{aligned}
&\mathcal{L}_0\Phi^{\text{(NR)}}_{m, n}=(n+\hat{h})\Phi^{\text{(NR)}}_{m, n}\,, \\
&\mathcal{W}_0\Phi^{\text{(NR)}}_{m, n}=i m\Phi^{\text{(NR)}}_{m, n}\,, \\
&\mathcal{L}_+\Phi^{\text{(NR)}}_{m, n}\approx n\Phi^{\text{(NR)}}_{m, n-1}\,, \\
&\mathcal{L}_-\Phi^{\text{(NR)}}_{m, n} \approx 2im\hat{\Omega}\Phi^{\text{(NR)}}_{m, n+1}\,.
\label{act}
\end{aligned}
\end{equation}
In the last two equations, we have expanded $\d x$ in $\mathcal{L}_{\pm}$ and used the eikonal conditions $|m|\gg 1$, $ |\hat{\omega}_R| \delta x^2 \gg 1$. Therefore, the $sl(2,\md{R})_{\text{ISO}}$ is retained in the solution space of near-ring eikonal QNMs.

In addition to the retained isometry, in the near-ring region the equation of motion Eq.~\eqref{EOM1} has an emergent conformal symmetry. To see this, we define the creation and annihilation operators as\footnote{Here, our definitions are slightly different from article  \cite{Kapec:2022dvc} for convenience.} $a_{\mp}$, 
\begin{equation}\label{aa}
\begin{aligned}
a_{-}&=\frac{e^{-T}}{\sqrt{2 k \hat{\omega}_R}}(i\partial_x+k \hat{\omega}_R \delta x)\,,\\
a_{+}&=\frac{e^{T}}{\sqrt{2k \hat{\omega}_R}}(\partial_x+ik\hat{\omega}_R \delta x)\,,
\end{aligned}
\end{equation}
which satisfy the Heisenberg algebra, $[a_{+},a_{-}]=1$. With the ansatz $\Phi = e^{im\p} e^{(-i\hat{\omega}_R+\hat{\omega}_I)T} \psi(x) $, the equation of motion in the near-ring region  becomes 
\be
\label{EOM2}
\bigg(a_-a_++\f{1}{2}\bigg) \psi(x) = -\hat{\omega}_{I} \psi(x) \,.
\ee
Hence, the solution space of eikonal QNMs in near-ring region  is in fact a Fock space. There is a single vacuum annihilated by $a_{+}$, which is exactly the fundamental mode in Eq.~\eqref{nearring},
\begin{equation}
\label{0}
\Phi_{m,0}^{\text{(NR)}} \propto e^{-\frac{i}{2}k\hat{\omega}_R \delta x^2},\hspace{3ex}\hat{\omega}_I=-\frac{1}{2}\, , 
\end{equation}
and the descendants are obtained by acting the creation operators on the vacuum, $\Phi_{m,n}^{\text{(NR)}} \propto a_-^n \Phi_{m,0}^{\text{(NR)}}$. The highest-weight condition ensures that the successive states are non-divergent, and do not have any poles in the near-ring region. Noting that
\begin{equation}
\label{ap}
a_{+} \left(\delta x^n e^{-\frac{i}{2}m \hat{\Omega}k \delta x^2} \right)= \frac{e^{T}}{\sqrt{2k \hat{\omega}_R}}n \delta x^{n-1} e^{-\frac{i}{2}m \hat{\Omega}k \delta x^2}\,
\end{equation}
is consistent with the third equation in Eq.~\eqref{act}, up to a overall factor, we have the relation
\begin{equation}\label{eq1118}
\mathcal{L}_{+} \sim L_+=\hat{s} a_{+}\,,
\end{equation}
where $\hat{s}$ is a constant to be determined. Comparing Eq.~\eqref{ap} with the action of $\mathcal{L}_+$ in Eq.~\eqref{act}, we have
\be
\mathcal{L}_+\Phi^{\text{(NR)}}_{m, n} \approx n\Phi^{\text{(NR)}}_{m, n-1} \hspace{2ex}\sim \hspace{2ex} L_+\Phi^{\text{(NR)}}_{m, n} = -\f{n \hat{s}}{\sinh{\td{x}}\sqrt{2k \hat{\omega}_R}}\Phi^{\text{(NR)}}_{m, n-1}\,,
\ee
where we have used  Eq.~\eqref{nearring}. Thus, we can determine that
\be\label{ss}
\hat{s} \approx -\sinh{\td{x}}\sqrt{2k \hat{\omega}_R}\,.
\ee
Based on the above analysis, we would like to propose a new construction of $sl(2,\md{R})$ algebra by using the creation and annihilation operators. 

In \cite{Hadar:2022xag}, the authors have proposed a construction of $sl(2,\md{R})$ which involved two highest-weight representations with the weights $1/4$ and $3/4$, respectively. These two towers of modes must be interwoven to form the entire QNMs. We aim to solve this perplexing ``two towers'' problem in the following construction. Firstly, we define $\mathcal{N}(L_{\pm})$ to be the number of $a_-$ minus the number of $a_+$ in $L_{\pm}$. Then, we require 
\be\label{NN}
\mathcal{N}(L_-) = +1\,, \quad \mathcal{N}(L_+) = -1\,.
\ee
To form an $sl(2,\md{R})$ algebra, $L_0$ should be of the form $a_{-}a_{+}+\hat{h}$ so that
\be
[L_0,L_p]=-p L_{p}\,, \qquad (p=+,-)\,.
\ee
Note that we add $\hat{h}$ to the expression of $L_0$ to give the correct highest weight. Then, by virtue of the Heisenberg algebra, the generators satisfying Eq.~\eqref{NN} can always be written as
\begin{equation}
\begin{aligned}
\label{FG}
L_{+}&=F(a_{-} a_{+})a_{+}\,,\\
L_0&=a_{-} a_{+}+\hat{h}\,,\\
L_{-}&=a_{-}G(a_{-} a_{+})\,,
\end{aligned}
\end{equation}
where $F(a_{-} a_{+}),G(a_{-} a_{+})$ are two functions constrained by
\begin{equation}\label{eq121213}
[L_{+},L_{-}]=2 L_0.
\end{equation}
Acting both sides of Eq.(\ref{eq121213}) on the state $\Phi^{(\text{NR})}_{mn}$, we obtain
\begin{equation}
(n+1)F(n)G(n)-n F(n-1)G(n-1)=2(n+\hat{h})\,.
\end{equation}
The general solution of the difference equation for $n F(n-1)G(n-1)$ is
\begin{equation}
n F(n-1)G(n-1)=n^2-n+2\hat{h}n+C\,,
\end{equation}
where $C$ is a constant. The boundary condition at $n=0$ gives $C=0$. Thus, the two functions $F$ and $G$ satisfy
\be\label{FG2}
F(a_{-} a_{+})G(a_{-} a_{+})=a_{-} a_{+}+2\hat{h}\,.
\ee
Therefore, the generators Eq.~\eqref{FG} form an $sl(2,\md{R})$ algebra which we denote as $sl(2,\md{R})_{\text{QNM}}$. So far, the $sl(2,\md{R})_{\text{QNM}}$ algebra is totally constructed from the Heisenberg algebra regardless of the physical situation. However, we expect $L_{\pm}$ to raise and lower the overtone numbers of $\Phi^{\text{(NR)}}_{mn}$, in the same way as $sl(2,\md{R})_{\text{ISO}}$. Comparing Eq.~\eqref{act} with Eq.~\eqref{eq1118}, we set
\be
F(x)=\hat{s}\,,  \, \
G(x)=\hat{s}^{-1}(x+2\hat{h})\,.
\ee
Therefore, we obtain an $sl(2,\md{R})_{\text{QNM}}$ algebra  which is generated by
\begin{equation}
\begin{aligned}
\label{eeqq24}
& L_{+}=\hat{s}a_{+}\,, \\
& L_0=a_{-} a_{+}+\hat{h}\,, \\
& L_{-}=\hat{s}^{-1}(a_{-}^2 a_{+}+2\hat{h}a_{-})\,.
\end{aligned}
\end{equation}
Acting $L_0$ and $L_-$ to the QNMs in Eq.~\eqref{nearring}, we find
\begin{equation}
\begin{aligned}
L_0\Phi^{\text{(NR)}}_{mn} & = (n+\hat{h})\Phi^{\text{(NR)}}_{mn} = \mathcal{L}_{0}\Phi^{\text{(NR)}}_{mn}, \\
L_-\Phi^{\text{(NR)}}_{mn} & = \hat{s}^{-1} \f{2m\hat{\Omega}k}{\sqrt{2k \hat{\omega}_R}} (-\sinh{\td{x}}) (a_{-} a_{+}+2\hat{h}-1) \Phi^{\text{(NR)}}_{mn+1} \nn \\
&\approx (2\hat{h}+n)\Phi^{\text{(NR)}}_{mn+1} \approx 2im\hat{\Omega}\Phi^{\text{(NR)}}_{mn+1} = \mathcal{L}_-\Phi^{\text{(NR)}}_{mn} .
\end{aligned}
\end{equation}
Hence, in our construction, the isometry and the emergent conformal symmetry in the near-ring region are completely equivalent when acting on the eikonal QNMs
\be
sl(2,\md{R})_{\text{ISO}}\Phi^{\text{(NR)}}_{mn} \simeq sl(2,\md{R})_{\text{QNM}}\Phi^{\text{(NR)}}_{mn}\, .
\ee
In conclusion, the solution space of near-ring eikonal QNMs forms one highest-weight representation
\begin{equation}\label{slqnm1}
\Phi^{\text{(NR)}}_{m, n}\approx \left(\frac{1}{2 i m \hat{\Omega}}\right)^n L^n_{-} \Phi^{\text{(NR)}}_{m, 0}\, ,  \quad n = 0,1,2...
\end{equation}
which satisfy
\be
L_0\Phi^{\text{(NR)}}_{m, n}=(n+\hat{h})\Phi^{\text{(NR)}}_{m, n}\,, \quad \quad
L_+\Phi^{\text{(NR)}}_{m, n}\approx n\Phi^{\text{(NR)}}_{m, n-1}\,.
\ee
We want to stress that the key point in the construction is creating the highest-weight representation from a single mode Eq.~\eqref{0}.

\section{Correspondence between QNMs and PR phase space }
In \cite{Kapec:2022dvc}, by noticing that the $sl(2,\md{R})_{\text{PR}}$ construction of Eq.~\eqref{choose1} is a simple deformation of $sl(2,\md{R})_{\text{ISO}}$, the authors expected to connect $sl(2,\md{R})_{\text{PR}}$ to $sl(2,\md{R})_{\text{QNM}}$ via the isometric group. 
In this section, we propose a direct connection between the sub-RPS of the photon ring and the coherent state space of $sl(2,\md{R})_{\text{QNM}}$.  The  sub-RPS is a two-dimensional space labeled by $(H_+, H_0, H_-)$ with the constraint $H_0^2=H_{+} H_{-}$, and the coherent state space is also a two-dimensional space labeled by $|A,\theta\rangle$ which will be introduced below. 

In the solution space of near-ring QNMs spanned by $\{\Phi^{(\text{NR})}_{m,n}\}$ in Eq.~\eqref{slqnm1}, for convenience, we denote $\Phi^{(\text{NR})}_{m,0}$ as $|0\rangle$ where we have suppressed the angular quantum number. With the help of the Baker-Campbell-Hausdorff (BCH) formula, a generic element of $sl(2,\md{R})_{\text{QNM}}$ group can be rewritten as\footnote{This must be true in the neighborhood of the identity operator.} $e^{\theta L_{-}} e^{A L_0} e^{B L_{+}}$.\label{ass} Such an element acting on the vacuum leads to 
\begin{equation}
\label{bch}
|A,\theta\rangle=e^{\theta L_{-}} e^{A L_0} e^{B L_{+}}|0\rangle=e^{\hat{h} A+\theta L_{-}}|0\rangle\,.
\end{equation}
The coherent state space of $sl(2,\md{R})_{\text{QNM}}$ is defined to be composed of all the $|A,\theta\rangle$ states. Then, the transformation of $|A,\theta\rangle$ under the action of $sl(2,\md{R})_{\text{QNM}}$ is
\begin{equation}\label{tran1}
\begin{aligned}
e^{\alpha L_{+}}|A,\theta\rangle&=\left|A-2\ln|1-\alpha\theta|,\frac{\theta}{1-\alpha\theta}\right\rangle,\\
e^{\alpha L_{0}}|A,\theta\rangle&=\left|A+\alpha,e^\alpha \theta\right\rangle,\\
e^{\alpha L_{-}}|A,\theta\rangle&=\left|A,\theta+\alpha\right\rangle.
\end{aligned}
\end{equation}
Therefore, the coherent state space $\{|A,\theta\rangle\}$ also forms a representation of $sl(2,\md{R})_{\text{QNM}}$. 

Furthermore, the actions of $sl(2,\md{R})_{\text{PR}}$ on the  sub-RPS $(H_+, H_0, H_-)$ can give a similar transformation. By defining $A' = -\ln{H_-}$, $\t' =H_0/H_-$, we find that as long as the phase space satisfies the constraint $H_0^2 = H_+ H_-$, the following equations hold
\begin{equation}\label{tran2}
\begin{aligned}
e^{- \alpha H_+} \t' e^{\alpha H_+} & = 
\frac{\t'}{1 - \alpha \t'}\,, \\
e^{- \alpha H_0} \t' e^{\alpha H_0} & =  e^{\alpha} \t'\,, \\
e^{- \alpha H_-} \t' e^{\alpha H_-} & =  \t' +\alpha\,, \\
e^{- \alpha H_+} A' e^{\alpha H_+} & =  A'  - 2 \ln \left(
1 - \alpha  \t'  \right)\,,\\
e^{- \alpha H_0}A'  e^{\alpha H_0} & = A'   + \alpha\,,\\
e^{- \alpha H_-} A'  e^{\alpha H_-} & =  A' \,. 
\end{aligned}
\end{equation}
These equations describe the sub-RPS flow under $H_{+}$, $H_{-}$ and $H_{0}$, which is consistent with Eq. (B.7-B.9)  in \cite{Kapec:2022dvc}. Comparing (\ref{tran1}) with (\ref{tran2}), we find that the parameters $(A,\theta)$ and $(A',\t')$ transform exactly in the same way under $sl(2,\md{R})_{\text{QNM}}$ and $sl(2,\md{R})_{\text{PR}}$, respectively, with $\alpha$ being the transformation parameter varying along the flow. Hence, it is natural to propose that the coherent states of $sl(2,\md{R})_{\text{QNM}}$ exactly correspond to the points in the sub-RPS. More precisely, a coherent state denoted by $(A,\theta)$ and a  point $\left({H_0},{H_{-}}\right)$ in the sub-RPS can be related by
\bea
A\sim-\ln H_{-}\, , \quad
\theta\sim\frac{H_0}{H_{-}}\, ,
\eea
or equivalently,
\bea
H_0 \sim \t e^{-A}\, , \quad
H_- \sim e^{-A}\, . 
\eea
The correspondence above can be concluded by a formal expression, 
\be
\mathcal{T}|A,\theta\rangle = |T[A],T[\theta]\rangle\,,
\ee
where $\mathcal{T}$ stands for any operator of $sl(2,\md{R})_{\text{QNM}}$ and the $T[\cdot]$ is the associated transformation on the parameters $(A,\theta)$ given by (\ref{tran2}).

Since the correspondence between these two different systems has been built, we can further consider what happens to a curve in the sub-RPS. A general curve in the sub-RPS can be described by
\begin{equation}
f(H_0,H_{-})=0\, .
\end{equation}
We can use the correspondence to obtain a curve in the coherent state space
\begin{equation}
\label{c}
\tilde{f}(A,\theta)=f(\t e^{-A}, e^{-A})=0\,.
\end{equation}
The curve defines a state as
\begin{equation}\label{corres}
|\psi\rangle\equiv\int e^{-A} d A \wedge d \theta\, \delta\left(\tilde{f}(A,\theta)\right)|A,\theta\rangle\,.
\end{equation}
It can be checked that $e^{-A}d A \wedge d \theta$ is $sl(2,\md{R})$ invariant, i.e., 
\begin{equation}
e^{-T[A]} d T[A] \wedge d T[\theta]=e^{-A} d A \wedge d \theta\,.
\end{equation}
Then a curve $\tilde{f}(A,\theta)=0$ transforms to a curve described by $\tilde{f}(T^{-1}[A],T^{-1}[\theta])=0$, which in turn defines a state as
\begin{equation}
\begin{aligned}
T[|\psi\rangle]\equiv&\int e^{-A} d A \wedge d \theta\, \delta\left(\tilde{f}(T^{-1}[A],T^{-1}[\theta])\right)|A,\theta\rangle\\
=&\int e^{-T[A]} d T[A] \wedge d T[\theta] \,\delta\left(\tilde{f}(A,\theta)\right)|T[A],T[\theta]\rangle\\
=&\int e^{-A} d A \wedge d \theta\, \delta\left(\tilde{f}(A,\theta)\right)\mathcal{T}|A,\theta\rangle\\
=&\mathcal{T}|\psi\rangle\,.
\end{aligned}
\end{equation}
In this sense, the curves in the sub-RPS transform in exactly the same way as the state $|\psi\rangle$  which is a superposition of QNMs. Thus Eq.(\ref{corres}) establish an exact correspondence between the curves in the sub-RPS and the coherent state space of QNMs. With the help of Eq.~\eqref{corres}, we can further investigate the corresponding state with a null geodesic.

It is remarkable that the correspondence shown in this section is simply a consequence of the $sl(2,\md{R})$ algebra, not depending on any specific construction like in Eq.~\eqref{prconstru1}, Eq.~\eqref{prconstru2} or Eq.~\eqref{FG}, as the derivation of Eq.~\eqref{tran1} and Eq.~\eqref{tran2} only involves the commutation relations of the $sl(2,\md{R})$ algebra and an extra constraint Eq.~\eqref{constraint1} in the sub-RPS. Furthermore, in this correspondence, the coherent state space of $sl(2,\md{R})_{\text{QNM}}$ can even be replaced by the coherent state space of $sl(2,\md{R})_{\text{ISO}}$.

\section{Discussions}

In this note, we discuss the realizations of the emergent conformal symmetries near the photon ring of a warped black hole. There are three types of conformal symmetries\cite{Kapec:2022dvc}: the isometry $sl(2,\md{R})_{\text{ISO}}$, the emergent one $sl(2,\md{R})_{\text{PR}}$ in the phase space of the photon ring, and the other emergent one $sl(2,\md{R})_{\text{QNM}}$ in the solution space of high-frequency  scattering in the near region. We managed to find a new construction of $sl(2,\md{R})_{\text{QNM}}$, which provides more transparent relations with $sl(2,\md{R})_{\text{ISO}}$ and $sl(2,\md{R})_{\text{PR}}$. We showed that when acting the newly constructed $sl(2,\md{R})_{\text{QNM}}$ and $sl(2,\md{R})_{\text{ISO}}$ on the eikonal QNMs, we have the same results at the leading order, even though the two  constructions are different. Consequently, the entire eikonal QNMs fall into the highest-weight representations of $sl(2,\md{R})_{\text{QNM}}$ or equivalently of $sl(2,\md{R})_{\text{ISO}}$. We also showed that there is a correspondence between the   coherent state space of $sl(2,\md{R})_{\text{QNM}}$ and the phase space of the photon ring. 

In our study, we focus on the warped spacetime in three dimensions. As discussed in the article\cite{Hadar:2022xag},  the near-ring QNMs have emergent conformal symmetry in both the Schwarzschild and the Kerr black holes. The novel construction of the  $sl(2,\md{R})_{\text{QNM}}$ is algebraic and can be applied to the four-dimensional black hole as well.  

For the Schwarzschild black holes, the null geodesics are labeled by a single impact parameter, $\lambda = L/H$. Thus, Eq.~\eqref{choose1} and Eq.~\eqref{choose2} can be generalized to Schwarzschild black holes to construct $sl(2,\md{R})_{\text{PR}}$ algebra. These two schemes provide the same attraction flows toward the bound photon  orbit. Moreover, there exists a correspondence between the coherent state space and the photon phase space for the Schwarzschild black hole.

However, for the Kerr black holes, two impact parameters are involved in the photon ring, which are $\lambda = L/H$, $\eta = Q/H^2$ with $Q$ the Carter constant. Hence, the construction of $sl(2,\md{R})_{\text{PR}}$ in Kerr black holes necessarily requires the participation of the canonical pair $(T,H)$, which gives the right attraction flows towards the bound photon orbits, as shown in Fig.2 in \cite{Hadar:2022xag}. Besides, as the photon phase space in Kerr is of six dimension, how to establish the correspondence between the coherent state space and the phase space is an interesting question.


\section*{Acknowledgments}
The work is in part supported by NSFC Grant  No. 12275004, 11735001.

\bibliographystyle{utphys}
\bibliography{refs}
\end{document}